\documentclass[12pt]{article}
\usepackage{amssymb,pifont}
\usepackage{epsfig}
\usepackage{axofigs}

\def \bra {\langle}
\def \ket {\rangle}
\def \beq {\begin{equation}}
\def \eeq {\end{equation}}
\def \barr {\begin{array}{lc}}
\def \earr {\end{array}}
\def \comma {\; \; ,\, \;}
\def \ket {\rangle}
\def \up  {| \! \uparrow \ket}
\def \down {| \! \downarrow \ket}
\def \ud  {| \! \uparrow \downarrow \ket}
\def \lf  {\left (}
\def \rt  {\right )}
\def \pl  {\partial}
\def \d   {\delta}
\def \eps {\epsilon}
\def \sg  {\sigma}
\def \gm  {\gamma}
\def \Gm  {\Gamma}
\def \ZZ  {\mathbb{Z}_2}
\def \scri {\ensuremath{\mathcal{I}}}
\def \scriplus {\ensuremath{\mathcal{I}^+}}
\def \scriminus {\ensuremath{\mathcal{I}^-}}
\def \ZZ  {\mathbb{Z}_2}

\newcommand{\cdott}{\!\cdot\!}

\begin{document}

\begin{titlepage}

\begin{center}

{\hbox to \hsize{\hfill CU-TP-1122}}
{\hbox to \hsize{\hfill ITFA-2004-49}}

\bigskip

\vspace{3\baselineskip}

{\Large \bf De Sitter Holography with a Finite Number of States}

\bigskip
\bigskip
\bigskip

{\large \sc Maulik~Parikh\footnote{\tt mkp@phys.columbia.edu} {\rm and}
  Erik~Verlinde\footnote{\tt erikv@science.uva.nl}\\[1cm]
}

{\it $^1$ Department of Physics, Columbia University, New York, NY
  10027}\\[3mm]

{\it $^1$ Physics Department, University of Amsterdam, The Netherlands}\\[3mm]

\vspace*{1.5cm}

\large{
{\bf Abstract}\\
}
\end{center}
\noindent
We investigate the possibility that, in a combined theory of  quantum
mechanics and gravity, de Sitter space
is described by finitely many states. The notion of observer
complementarity, which states that  each observer has complete but
complementary information, implies that, for a single observer, the complete
Hilbert space describes one side of the horizon.  Observer
complementarity is implemented by identifying antipodal states with outgoing
states. The de Sitter group acts on S-matrix elements.  Despite the fact 
that the de Sitter group has no nontrivial
finite-dimensional unitary representations, we show that it is
possible to construct an S-matrix that is
finite-dimensional, unitary, and de Sitter-invariant. 
We present a class of examples that realize this idea holographically in terms of
spinor fields on the boundary sphere. The finite dimensionality is due to
Fermi statistics and an `exclusion principle' that truncates the
orthonormal basis in which the spinor fields can be expanded.

\end{titlepage}

\newpage

\setcounter{page}{2}
\setcounter{footnote}{0}

\tableofcontents

\bigskip
\bigskip

\section{Introduction}

Suppose we had the holographic formulation of quantum gravity or
string theory in asymptotic de Sitter space -- how would we recognize
it as such? Experience with Matrix theory and AdS/CFT suggests that
the answer is: through the symmetries of de Sitter space. In all known
examples of holography, bulk isometries appear as symmetries of the
boundary theory. For example, in AdS/CFT, the presence of
five-dimensional anti-de Sitter space can be deduced from the dual
theory's $O(2,4)$ conformal symmetry. 
This idea has been used to propose that $d$-dimensional
de Sitter space is holographically dual to a $d-1$ dimensional 
boundary theory which has the $O(1,d)$ de Sitter isometry group as a 
symmetry \cite{dscft}. But, unlike AdS, de Sitter space has cosmological
horizons. These horizons have finite area, and hence lead to a finite
de Sitter entropy.  
There are various interpretations of the finiteness of the de Sitter 
entropy, but one possibility is that it indicates that the Hilbert
space of the holographic dual 
has only a finite number of states \cite{banks,fischler}. We will
consider this strong version of the holographic principle here. 

Now, there is an apparent tension between the de Sitter group and
the finiteness of the number of states. This is
because the de Sitter group is noncompact, and there is an old theorem
that says that only compact groups have nontrivial finite-dimensional 
representations that are unitary. If the Hilbert space is finite, it
cannot furnish a nontrivial
unitary representation of the de Sitter group. There have thus been
claims that de Sitter space has no holographic dual \cite{gks}, or that
the symmetry group is not the de Sitter group \cite{witten}.
 
Nevertheless, in this paper we will try to have our cake and eat it
too. Namely, we will consider the possibility that there is a 
holographic description of de Sitter space that is unitary, respects
the symmetries of de
Sitter space, and yet has only a finite number of states.
Our description is motivated by the principle of observer
complementarity \cite{dSZ2,stu,thooft,compl}. 
One formulation of this principle asserts that every single observer 
has complete information; there is no need to consider states or events
that live or happen outside the observer horizon. For a given observer
the Hilbert space  
consists only of states that are accessible to him or her; these states
are sufficient to describe  
all events that take place inside the observer horizon. 
The measurable 
observables are the S-matrix elements that give the probabilities for
these events. For unitarity it is then sufficient that the Hilbert
space contains states with positive norm, and that the S-matrix be
unitary. The clash between unitarity
and the finiteness of the number of states is removed because the
Hilbert space associated to one observer does not have to be a
representation  of the full de Sitter group, but only of the compact
subgroup that leaves the horizon invariant. 

However, different observers are 
related by the de Sitter group
and, in some sense, should be equivalent. How, then, should we
implement de Sitter symmetry?
We propose that the de
Sitter group, even though it does not act on 
individual states, transforms the 
elements of the S-matrix. 
De Sitter invariance implies that the S-matrix respects the symmetries of de 
Sitter space, just as the S-matrix in Minkowski space respects
Poincar\'e symmetries.
In other words, the tensor product of in and out Hilbert spaces should
form a representation of the de Sitter group in such a way that the
tensor product state corresponding 
to the S-matrix is invariant. The in and out Hilbert spaces do not
by themselves form representations of the de Sitter group because they
only form representations of the group that preserves the horizon.
The requirement of de Sitter invariance of the tensor product states
relates different S-matrix elements and leads to selection rules that
reflect the underlying de Sitter symmetry.

The aim of this paper is to present a concrete realization of these
ideas by constructing 
an infinite class of finite-dimensional Fock spaces based on spinor
representations of the de Sitter group. 
This paper is organized as follows. After a brief review of the de Sitter
group in section 2, we present the main idea behind the construction
in section 3. 
The finitely many Fock states corresponding to a static
patch of de Sitter space do not form representations of the de Sitter group
but only of the (compact) rotation subgroup that preserves the observer's
horizon. Global de Sitter space is described by a tensor product of
two copies of the Fock space, one corresponding to each of an
antipodal pair of observers. The key point is that there are states in
the tensor product space that are de Sitter-invariant. In section 4,
we make the abstract discussion concrete by presenting two constructions of a 
finite-dimensional Fock space. 
In the
first construction, the dual theory is based on Dirac spinors. This is
perhaps the simplest toy model of de Sitter space
\cite{toystory}. A more general construction, presented in section 5,
elevates the Dirac spinors to a spinor field theory living on a
sphere. It is manifestly holographic in that the sphere can be thought
of as the boundary of de Sitter space. Moreover, it allows
representations that, while finite,
can be arbitrarily large. In section 6, we
identify the de Sitter-invariant states in the tensor product of the
antipodal Fock spaces. These singlet states are the key
to implementing the de Sitter symmetry. We
argue that, by making an antipodal identification of de Sitter space,
the two Fock spaces can be thought of as the space of initial and final
states. The invariant states can then be reinterpreted as de
Sitter-invariant S-matrix elements. Section 7 is a brief illustration of
how observer complementarity works in practice.

We mention here that an alternate approach with the same goal in mind
would be to q-deform the de Sitter group
\cite{albertodavid,lowe}. In particular, there exist finite and unitary
principal series representations of the cyclically q-deformed de
Sitter group.

\section{The de Sitter Group and the Little Group}

Let us briefly review the symmetries of de Sitter space
\cite{leshouches,klemm}. De Sitter space can be represented as a
timelike hyperboloid embedded in Minkowski space:
\begin{equation}
-{X^0}^2 + {X^1}^2 \dots + {X^d}^2 = +R^2 \; , \label{embed}
\end{equation}
where $X^I$ are Cartesian embedding coordinates. In this form, the
$O(1,d)$ isometry group of $d$-dimensional de Sitter space is
manifest; the $d$-dimensional de Sitter group is
therefore the Lorentz group in $d+1$ dimensions. The Lorentz generators
are
\begin{equation}
M_{IJ} = -i (X_I \pl_J - X_J \pl_I ) \; .
\end{equation}

Now consider a geodesic observer in de Sitter space. Such an observer
moves along a worldline which, in the Minkowski embedding space, is
traced by a trajectory of constant acceleration. This worldline
is generated by boosts. Without loss of generality, let
the observer stay at the north pole, which we take to be in the
positive $X^d$ direction. Then the Hamiltonian is
\begin{equation} 
H = i \lf X^0 {\pl \over \pl {X^d}} + X^d {\pl \over \pl X^0} \rt \; .
\end{equation} 
The generators that leave the observer's worldline invariant are the
rotations about the axis connecting the poles as well as the time
translation operator. Together these form the observer's little group
$O(d-1) \times R$. Note that since the rotation group $O(d-1)$ is
compact, it admits finite-dimensional unitary representations.

Indeed, this group appears also for a nongeodesic observer. The past
and future horizons of an observer are completely determined by
specifying a point on the sphere at $\scriminus$ and another on the
sphere at $\scriplus$. We can just consider two points on a single
sphere. Then the generators that correspond to rotations in Minkowski
space move both these points while keeping their angular separation
fixed. The boosts also move the points but change their angular
separation. By means of de Sitter symmetries we can set the points to
be the two poles of the sphere. On the sphere, the de Sitter group
acts as the Euclidean conformal group $O(1,d)$. It is clear that the
transformations that leave the poles invariant consist of the
rotations about the axis formed by the poles, as well as the dilation
operation. Hence the group that preserves the horizon is just
$O(d-1) \times R$. 

As the rotation group $O(d-1)$ will play a key role in what follows,
it is convenient to relabel the de Sitter algebra in terms of indices
$i, j$ that run from 1 to $d-1$:
\begin{equation}
J_{ij} \equiv M_{ij} \qquad P_i \equiv M_{di} \qquad K_i \equiv M_{0i}
\qquad H \equiv M_{0d} \; .
\end{equation}
Here $J_{ij}$ generate the rotation group $SO(d-1)$, $P_i$ are
momentum operators, $K_i$ are the boosts, and $H$ is the generator of
time translations. Then, in terms of these generators, the de Sitter
algebra consists of the $so(d-1)$ rotation algebra,
\begin{equation}
{[J_{ij}, J_{km}]} = -i \lf \d_{jk} J_{im} - \d_{ik} J_{jm} - \d_{jm}
J_{ik} + \d_{im} J_{jk} \rt \; , \label{angmom}
\end{equation}
as well as
\begin{eqnarray}
{[J_{ij}, P_k]} \; \; =& -i \d_{jk} P_i + i \d_{ik} P_j \qquad \qquad
{[P_i, P_j]} &= +i J_{ij} 
\nonumber \\
{[J_{ij}, K_k]} \; \; =& -i \d_{jk} K_i + i \d_{ik} K_j \qquad \;
\; \; \; \, {[K_i, K_j]} &= -i J_{ij} 
\nonumber \\
{[K_i, H]} \; \; =&  +i P_i \qquad \qquad \qquad \qquad \qquad
\; \! \! \! \! \! {[P_i, H]} &= + i K_i 
\nonumber \\
{[P_i, K_j]} \; \; =& - i \d_{ij} H \qquad \qquad \qquad \qquad
{[J_{ij}, H]} &= 0 \; .
\end{eqnarray}
Note that $P_i$ and $J_{ij}$ together generate $SO(d)$. 

\section{Fock States and Tensor Product States}

The view we will take here is that the finiteness of the entropy of de
Sitter space implies a finite number of states in the holographically
dual theory. Now, since these states are associated with a horizon,
one might expect that they transform under representations of the
group that keeps the horizon fixed, namely $O(d-1) \times R$. This is
almost correct, but it must be remembered that this dual theory
represents (or rather, defines) a quantum theory of gravity for de
Sitter space. In quantum gravity, one keeps the geometry on the
boundaries fixed while allowing the bulk geometry to fluctuate. Only
those transformations that are defined at the boundary should be used
to label states. In particular, the Hamiltonian moves a point from the
boundary into the bulk and so is not a well-defined operation at the
boundary. By contrast, the rotations $O(d-1)$ are well-defined
operations at the boundary. So we conclude that the states in the
dual theory should transform not under the noncompact group $O(d-1)
\times R$ but only under the compact rotation group $O(d-1)$.

We shall therefore take an observer's one-particle Hilbert space,
$h_I$, to be a finite-dimensional unitary representation of
$O(d-1)$. We denote the corresponding Fock space by ${\cal H}_I$. In
order to obtain a finite-dimensional Fock space, the particles 
need to obey fermionic statistics so that there are at most a finite
number of particles. Similarly, the states accessible to the antipodal
observer are contained in a finite-dimensional Fock space ${\cal
H}_{II}$ constructed out of a one-particle Hilbert space, $h_{II}$,
that is isomorphic to that of the first observer. Later we will see
that under the antipodal identification the two Hilbert spaces can be
interpreted as in and out Hilbert spaces.

\begin{figure}[hbtp]
        \centering
\begin{picture}(0,140)
\EBox(-50,20)(50,120)
\DashLine(-50,20)(50,120){3}
\DashLine(-50,120)(50,20){3}
\Text(0,125)[b]{$\mathcal{I}^+$}
\Text(0,15)[t]{$\mathcal{I}^-$}
\Text(-52,70)[r]{N}
\Text(52,70)[l]{S}
\Curve{(0,70)(25,50)(50,40)}
\Curve{(0,70)(25,62)(50,60)}
\Curve{(0,70)(25,78)(50,80)}
\Curve{(0,70)(25,90)(50,100)}
\DashCurve{(-50,40)(-25,50)(0,70)}{1}
\DashCurve{(-50,60)(-25,62)(0,70)}{1}
\DashCurve{(-50,80)(-25,78)(0,70)}{1}
\DashCurve{(-50,100)(-25,90)(0,70)}{1}
\Text(30,70)[]{$I$}
\Text(-30,70)[]{$II$}
\end{picture}
        \caption[]{\small Penrose diagram of de Sitter space. Region $I~(II)$
          corresponds to the static patch of an observer on the south (north)
          pole. The solid lines indicate equal time slices in the static time,
          they are Cauchy surfaces for region $I$. The dotted lines are their
          antipodal images, and constitute Cauchy surfaces for region $II$.
          When a solid line is continued through the horizon, onto its
          antipodal image, it constitutes a Cauchy surface for the whole
          space. The global state is defined on a Cauchy surface for
          the whole space and therefore lives in the tensor product of
          the Fock spaces of regions $I$ and $II$.}
      \label{cauchy}
\end{figure}
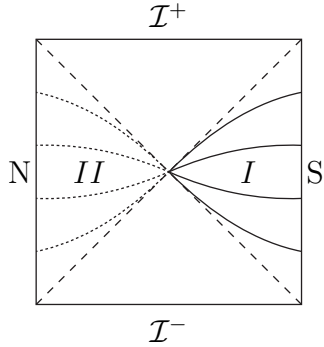

The full global state specifying the state of global de Sitter space
is a tensor product of states in the Fock spaces of antipodal pairs of
observers; see Fig. (\ref{cauchy}). The tensor product states live in
the direct product space ${\cal H}_I \otimes {\cal H}_{II}$, which we
denote ${\cal H}$:
\begin{equation}
{\cal H} = {\cal H}_I \otimes {\cal H}_{II} \; .
\end{equation}
A typical basis state of the dual to global de Sitter space is therefore
\begin{equation}
|m \ket_I \otimes |n \ket_{II} \; ,
\end{equation}
where $m$ and $n$ are quantum numbers of (the Cartan subalgebra of)
$O(d-1)$. The general state can be written as 
\begin{equation}
|\Psi \ket = \sum_{m, n} C_{mn} |m \ket_I \otimes |n \ket_{II} \;
 . \label{psi}
\end{equation}
Mixed states that cannot be written in the form
$|\psi \ket_I \otimes |\phi \ket_{II}$ are to be interpreted as states
that correlate the antipodal observers. The state corresponding to a
particle beyond the future horizons of both the observers might be an
example of such a mixed state.

Since the global Fock space is a direct product of the individual
Fock spaces, it follows that the global one-particle Hilbert
space must be a direct sum of the one-particle Hilbert spaces of the
pair of antipodal observers. One can check this by counting the number
of states: if there are $n$ states in a single observer's Hilbert
space, then there are $2^n$ states in the Fock space, and hence $2^n
\times 2^n = 2^{2n}$ in the global (tensor product) Fock space. So the
global Hilbert space, $h$, has $2n$ states, indicating that it is a
direct sum of the two Hilbert spaces:
\begin{equation}
h = h_I \oplus h_{II} \; . \label{directsum}
\end{equation}

Consider now a general de Sitter transformation, $M$. The Fock spaces
${\cal H}_I$ and ${\cal H}_{II}$ are $O(d-1)$-invariant by
construction, but they are not de Sitter-invariant. This is because a
de Sitter transformation that is not in $O(d-1) \times R$ will in
general move a particle out of the spacetime accessible to one
observer and into the spacetime accessible to the antipodal observer.
If we think of de Sitter transformations as acting actively, then they
simply move a point to some other point in de Sitter space. Such
transformations can take a particle out of an observer's horizon into
the antipodal observer's horizon. So on the one-particle Hilbert space
$M$ acts as
\begin{equation}
M: h_I \oplus h_{II} \to h_I \oplus h_{II} \; . \label{dSonHilbert}
\end{equation}
That is, while $h_I$ and $h_{II}$ form representations of $O(d-1)$,
their direct sum forms a representation of $O(1,d)$.
More generally, on the space of tensor products of Fock states, $M$ acts as
\begin{equation}
M: {\cal H}_I \otimes {\cal H}_{II} \to {\cal H}_I \otimes {\cal
 H}_{II} \; ,
\end{equation}
so again the direct product space forms a representation of $O(1,d)$
rather than just of $O(d-1)$. Of course, since the individual Fock
spaces are finite-dimensional, the tensor product space is also
finite-dimensional. Hence it also falls afoul of the theorem
forbidding unitary finite-dimensional representations of the de Sitter
group. But note that the theorem requires only that {\em nontrivial}
unitary representations of the de Sitter group be
infinite-dimensional. This provides the key loophole: we can find
certain special tensor product states that transform in a {\em
trivial} representation of the de Sitter group. That is, we will
look for those singlet tensor product states, $|\Psi_S \ket$, that are
invariant under the de Sitter group:
\begin{equation}
M |\Psi_S \ket = 0 \; .
\end{equation}
The Fock space accessible to an individual observer nevertheless has a
finite number of states, so the entropy is not zero. In section 6, we
will see that if we identify antipodal points in de Sitter space, the
tensor product states become amplitudes for processes; the singlet
tensor product states becomes the de Sitter-invariant
S-matrix for de Sitter space.

\section{Finite-Dimensional Hilbert Spaces}

In this section we will illustrate the above ideas by
constructing two concrete representations. One is a spinor
representation, which is perhaps the simplest toy model of a de Sitter
dual. This representation was presented in a companion paper
\cite{toystory}; we review it here. The second representation is more
manifestly holographic. It consists of fields that form
finite-dimensional representations of the conformal group of
the sphere at $\scri$.

\subsection{A spinor as a ``toy model'' of de Sitter space}
\label{diracspinors}

Let us start with the spinor representation. There are a couple of
motivations for considering spinors. Because of spin-statistics, the
Fock space of fermions can be finite-dimensional. For
bosons, the Fock space would be infinite-dimensional even for a
finite-dimensional Hilbert space as there is no restriction on the
number of bosons. Another reason for considering spinor
representations is that the number of components of a Dirac spinor doubles
when the dimension is increased by two. For a Dirac spinor
representation $R$,
\begin{equation}
{\rm dim}~ O(d-1)_R = {1 \over 2} {\rm dim}~ O(1,d)_R \; .
\end{equation}
This allows us to write a global Hilbert state, which transforms under
$O(1,d)$ as a direct sum of two states that transform under $O(d-1)$,
precisely as required by (\ref{dSonHilbert}).

The $\Gm$ matrices obey the Clifford algebra $\{\Gm_I, \Gm_J \} = 2
\eta_{IJ}$. We want to choose a representation which has the property
that the $O(d-1)$ subalgebra appears in the diagonal. A convenient
representation is
\begin{equation}
\Gm_i = \sg_3 \otimes \gm_i = \lf \barr \gm_i \; \; \; \; 0 \\ 0 \;
-\gm_i \earr \rt \comma \Gm_0 = i \sg_2 \otimes 1 = \lf \barr \; \; 0
\; \; 1 \\ 
-1 \; 0 \earr \rt \comma \Gm_d = \sg_1 \otimes 1 = \lf \barr 0 \; \; 1 \\ 1
\; \; 0 \earr \rt \; ,
\end{equation}
where $\gm_i$ are the gamma matrices for the Clifford algebra
$\{\gm_i, \gm_j\} = 2 \d_{ij}$, with $i, j$ running from 1 to $d-1$.

The de Sitter generators can then be written as
\begin{equation}
M_{IJ} =  -{i \over 4} [ \Gm_I , \Gm_J ] \; .
\end{equation}
To be specific, consider four-dimensional de Sitter space. The
four-dimensional de Sitter group is $O(1,4)$ and the observer's
little group is $O(3)$. Then, in
the above representation, the de Sitter generators are
\begin{equation}
J_i = {1 \over 2} \lf \barr
\sg_i \; \; 0 \\ 0 \; \; \sg_i \earr \rt \comma
P_i = {i \over 2} \lf \barr \; \; 0 \; \; \sg_i \\ -\sg_i \; 0 \earr \rt \comma
K_i = {i \over 2} \lf \barr 0 \; \; \sg_i \\ \sg_i \; \; 0 \earr \rt \comma
H = {i \over 2} \lf \barr -1 \; 0 \\ \; \; 0 \; \; 1 \earr \rt \; ,
\end{equation}
where $\sg_i$ are the Pauli matrices and $J_i = {1 \over 2} \eps_{ijk}
J^{jk}$. The one-particle Hilbert space, $h_I$, consists of just two
states: the two-component spinors, $\up$ and $\down$. The antipodal
observer has an isomorphic Hilbert space, $h_{II}$, and the above
$4 \times 4$ matrices act on the direct sum of these two Hilbert
spaces. Notice that $P_i$ and $K_i$ are off-diagonal, indicating that
they do not act within the Hilbert space of a single observer.

In this representation, $J_i$ and $P_i$ are hermitian whereas $K_i$
and $H$ are not. The fact that the Hamiltonian is not hermitian is not
a problem: recall that these are representations in a dual theory at
the boundary where energy is not a ``good quantum number.''
It is the nonhermiticity of the Hamiltonian that allows
us to evade the no-go theorem of \cite{gks}.
It is also convenient to define the ladder operators
\begin{equation}
L_{i+} \equiv - i P_i + i K_i = - \lf \barr 0 \; \; \; 0 \\ \sg_i \;
\; 0 \earr \rt \comma 
L_{i-} \equiv -i P_i - i K_i = \lf \barr 0 \; \; \sg_i \\ 0 \; 
\; 0 \earr \rt \; .
\end{equation}
Note that these obey an unusual hermiticity relation:
$L^\dagger_{i+} = - L_{i-}$.

\subsection{Finite conformal fields on the sphere}

The preceding construction works as a toy model for de Sitter space:
it is obviously finite-dimensional and, as we will see later, one can
find a de Sitter-invariant S-matrix that is unitary. However, the
model has two drawbacks. First, it has too few states to describe
semi-classical de Sitter space and second, the dual theory is not
manifestly holographic -- it is not obvious that it lives on the
boundary sphere at $\scri$. In this subsection, we study larger
representations of the conformal group of the sphere. The
eigenfunctions of the Cartan subalgebra of the conformal group will
turn out to be a finite set of polynomials; these form a finite
orthonormal basis in which to expand fields living on the boundary sphere.

The $d$-dimensional de Sitter group $O(1,d)$ is also the
$d-1$-dimensional Euclidean conformal group. In terms of the
generators we defined earlier, we have
\begin{equation}
L_{i\pm} = -i(P_i \pm K_i) \comma L_0 = iH \; .
\end{equation}
These obey the angular momentum algebra (\ref{angmom}), as well as the
$sl(2)$ algebra
\begin{equation}
{[L_{i\pm}, L_0]} = \pm L_{i\pm} \comma {[L_{i+}, L_{j-}]} = 2 (-i
J_{ij} + \d_{ij} L_0)
\end{equation}
and also
\begin{eqnarray}
{[J_{ij}, L_{k\pm}]} \; \; =& -i \d_{jk} L_{i\pm} + i \d_{ik} L_{j\pm}
\qquad \qquad
{[J_{ij},L_0]} &= 0
\nonumber \\
{[L_{i+}, L_{j+}]} \; \; =& 0 \; \qquad \qquad \qquad \qquad \qquad
\; \; \; {[L_{i-}, L_{j-}]} &= 0
\end{eqnarray}
A representation of the conformal algebra is
\begin{eqnarray}
J_{ij} & = & -i (x_i \pl_j - x_j \pl_i) \nonumber \\
L_{i-} & = & -\pl_i \nonumber \\
L_{i+} & = & x^2 \pl_i - 2 x_i (x^k \pl_k - \Delta) \nonumber \\
L_0 & = & -(x^k \pl_k - \Delta) \; . \label{confalgebra}
\end{eqnarray}
(Raised indices are the same as lowered and are only used to indicate
summation.)

To find the representation space, notice that $L_{i-}$ annihilates $1$
while $L_{i+}$ annihilates $(x^2)^\Delta$. Hence the representation
is bounded in both directions and we can generate the whole
representation by acting repeatedly on $(x^2)^\Delta$ with $L_{i-}$.
We find that a finite representation of the conformal group is spanned by
monomials of $x_i$. In particular, by acting with the
dilation operator, we get $L_0 ~ 1 = \Delta$ and $L_0 (x^2)^{\Delta}
= - \Delta$ i.e. the lowest weight state has conformal weight $\Delta$ while 
the highest weight state has {\em negative} conformal weight,
$-\Delta$. 

For example, when $\Delta = 1$, we obtain the $d+1$-dimensional representation:
\begin{equation}
1 \comma x_i \comma x^2 \; ,
\end{equation}
where $i$ ranges from 1 to $d-1$ as usual.
A conformal field $\phi(x)$ on the three-dimensional
sphere can be expanded in terms of the above basis,
\begin{equation}
\phi(x) = \phi_0 + \phi^i x_i + \phi_4 x^2 \; ,
\end{equation}
to provide a five-dimensional representation of the conformal
group. The five coefficients $\phi_M$ are just complex numbers. Here we
interpret the fact that all Taylor series terminate at a finite power
to be an indication of an underlying exclusion principle,
reminiscent of the ``stringy exclusion principle'' found in
AdS/CFT  \cite{MaldStrom}. Even though we do not have any obvious
noncommutativity, our description is also similar to the ideas of
describing de Sitter space  in terms of fuzzy spheres
\cite{banksfuzzy,miaoli}. 

Similarly, when $\Delta = 2$ we obtain the ${1 \over 2} d(d+3)$-dimensional
representation:
\begin{equation}
1 \comma x_i \comma x_i x_j \comma x^2 x_i \comma (x^2)^2 \; . \label{Delta=2}
\end{equation}
In general, a basis for a representation consists of all the
symmetric monomials whose order is less than or equal to $\Delta$, as
well as their inverses under
\begin{equation}
\phi(x) \to \phi'(x') = (x^2)^{\Delta} \phi(x') \; ,
\end{equation}
where $x_i \to {x_i}' = +x_i/x^2$. By commuting the field with $L_0$, one
can check that inversion takes a primary field of scaling dimension
$h$ to one with scaling dimension $-h$.

Thus far the discussion has been about the global Hilbert states since
we have been considering representations of the conformal group
$O(1,d)$. The Hilbert states accessible to a given observer, however,
transform as representations of $O(d-1)$. These 
states are labeled by $l$ and $m_k$ where $m_k$ are the eigenvalues of
the Cartan subalgebra of $so(d-1)$ (with $k$ running up to the rank of
$so(d-1)$) and $l(l+d-2)$ is the eigenvalue of the Casimir, $L^2$:
\begin{equation}
L^2 = (d-1) x \cdot \pl - x^2 \pl^2 + x^k x \cdot \pl \, \pl_k \; .
\end{equation}
To explicitly write down the coordinate representation of the Hilbert
states, we should take linear combinations of the monomials such that,
for $d=4$, they are eigenfunctions of $L^2$ and $L_3$. Denoting $\bra
x | l m \ket = \phi_{l,m}(x)$, we find that
\begin{equation}
\phi_{0,0} (x) =  N_0
\end{equation}
\begin{eqnarray}
\phi_{1,1} (x) & = & {N_1 \over \sqrt{2}} \lf x_1 + i x_2 \rt \nonumber \\
\phi_{1,0} (x) & = & -N_1 x_3 \nonumber \\
\phi_{1,-1} (x) & = & -{N_1\over \sqrt{2}} \lf x_1 - i x_2 \rt
\end{eqnarray}
\begin{eqnarray}
\phi_{2,2} (x) & = & N_2 (x_1 + i x_2)^2 \nonumber \\
\phi_{2,1} (x) & = & 2 N_2 (x_1 + i x_2)(-x_3) \nonumber \\
\phi_{2,0} (x) & = & \sqrt{6} N_2 \lf x^2_3 - {1\over 3} x^2 \rt \nonumber \\
\phi_{2,-1} (x) & = & 2 N_2( x_1 - i x_2)x_3 \nonumber \\
\phi_{2,-2} (x) & = & N_2 (x_1 - i x_2)^2 
\end{eqnarray}
Here the relative normalizations come from the $su(2)$ raising and lowering
operators, $L_{\pm} |l, m \ket = \sqrt{(l\pm m +1)(l \mp m)} |l, m \pm
1 \ket$. A common overall normalization, $N_i$, is still missing and depends
on a choice of inner product since that in turn defines the norm.

Finally, in order for our states to live not just in a vector space but
in a Hilbert space, we need to provide some additional
structure, namely a hermitian inner product. 
For a representation with highest weight $-\Delta$, a choice of
inner product is
\begin{equation}
(\phi_m, \phi_n) = \int {d^{d-1} x \over (1+x^2)^{d-1+\Delta}} \lf
  \phi^*_m(x) \phi'_n(x') + \phi_n(x) \phi^*_m(x') \rt \; .
\end{equation}
As a form, this is clearly hermitian and linear. Furthermore,
all the $|l, m \ket$ eigenstates are orthogonal and have positive
norm. Thus it is a true inner product and the 
vector space is therefore a Hilbert space.\footnote{An inner product
  space is a Hilbert space if it is Cauchy-complete with respect to
  the metric, $d(x,y)$, induced by the inner product, $d(x,y) = \bra x
  - y, x - y \ket$. Every finite-dimensional inner product space is
  complete so one need only verify the properties of the inner product.}
In addition, one can easily check that the inner product is
$O(d-1)$-invariant. We can also determine all the normalizations. For
example, for $d=4$ and $\Delta = 1$, the norm of the state $|1~0\ket$ is
\begin{equation}
(\phi_{1,0}, \phi_{1,0}) = 2 |N_1|^2 \int {d^3 x \over (1+x^2)^4}
  (-x_3) x^2 \lf {-x_3 \over x^2 }\rt = |N_1|^2 {\pi^2 \over 12} \; ,
\end{equation}
so that $N_1 = 2 \sqrt{3}/ \pi$ for $\Delta = 1$.

\section{Finite Conformal Spinors on the Sphere}

The preceding construction is not completely satisfactory, however, for two
reasons. First, the inner product says that $\phi(x)$ is dual to
$\phi'(x')$. However, there are states in the middle of the
representation (e.g. $x_i$ for $\Delta = 1$) that map to
themselves. Depending on the dimension and the representation, there
may even be an odd number of states. This is a problem because we
would like to be able to break up the global Hilbert state into a
direct sum of two Hilbert states, in accordance with $h = h_I \oplus
h_{II}$. A second problem is that, even though the one-particle
Hilbert space is finite-dimensional, the Fock space is still
infinite-dimensional since it contains states of the form $\phi^n
|{\rm vac} \ket$ with no restriction on the number of particles, $n$. 

Fortunately, we can solve both these problems by adding a spinor index
to the field, just as we did in section \ref{diracspinors}.
That is, we generalize the conformal algebra in (\ref{confalgebra}) to
\begin{eqnarray}
J_{ij} & = & -i (x_i \pl_j - x_j \pl_i) - {i \over 4} [\gm_i, \gm_j]
\nonumber \\
L_{i-} & = & -\pl_i 
\nonumber \\
L_{i+} & = & x^2 \pl_i - 2 x_i (x^k \pl_k - \Delta)
- {1 \over 2} x^k [\gm_i, \gm_k] 
\nonumber \\
L_0 & = & -(x^k \pl_k - \Delta) \; ,
\end{eqnarray}
where $\gm_i$ are the gamma matrices for the Clifford algebra,
$\{\gm_i,\gm_j\} = 2 \d_{ij}$.
We can check that, for $d=4$, if we write $J_i = {1 \over 2}
\eps_{ijk}J^{jk}$ then $J_i = L_i + S_i$, where $L_i$ is the orbital angular
momentum and $S_i = \sg_i /2$ is the spin operator.

\begin{figure}[hbtp]
 \centering
  \epsfysize=5 cm
  \hspace{2cm}\epsfbox{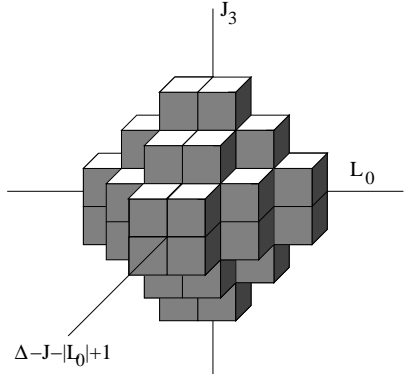}
 \centering
 \caption{A graphical depiction of the $\Delta=5/2$ representation in $d=4$ with 40 states.
The states are organized with respect to their $L_0$ eigenvalue, their
$J_3$ eigenvalue, and their total spin $J$. Both $L_0$ and $J_3$ take half
integer values between $-\Delta$ and $\Delta$, in this case $-5/2$ and
$5/2$. The total spin $J$ goes from $1/2$ till $\Delta=5/2$. We plot
$\Delta-J-|L_0|+1$ instead of $J$ since this gives a clearer picture.
{}From the diagram it is obvious how to extend it to representations with
other values of $\Delta$}
 \label{fig:reps}
\end{figure}

As before, $L_{i-}$ annihilates $1$, the highest weight state, so the
representation is bounded from below. The algebra has finite
representations if it is also bounded from above i.e. if the lowest
weight state is annihilated by $L_{i+}$. For a given $\Delta$, we find that 
\begin{equation}
L_{i+} (x^2)^{\Delta - 1/2} x \cdott \gm = 0 \; ,
\end{equation}
where $x \cdott \gm$ stands for $x_i \gm^i$. Hence we should take
representations with {\em half-integer} conformal dimension. The
monomials again form an eigenbasis for $L_0$. For example, we have
\begin{eqnarray}
\Delta = 1/2 & : & 1 \comma x \cdott \gm \; \nonumber \\
\Delta = 3/2 & : & 1 \comma x_i \comma x_i \,x \cdott \gm \comma x^2 \, x\cdott
\gm \nonumber \\
\Delta = 5/2 & : &
1 \comma x_i \comma x_i x_j \comma x_i x_j\,x \cdott \gm \comma
x^2\, x_i x \cdott \gm \comma x^4 \,x\cdott \gm \; . \label{fuzzyreps}
\end{eqnarray}
Each monomial has an $O(d-1)$ spinor index as well. 
For example, when $d=4$, these are the ${\bf 4}$, the ${\bf 16}$, and
the ${\bf 40}$ of $O(1,4)$ because each monomial comes with a two-component
spinor. In figure \ref{fig:reps} we have depicted the representation with 40 states.

In general, the rule is to take symmetric spinor-valued monomials up to
order $\Delta -1/2$ as well as their duals under 
\begin{equation}
\psi_a(x) \to \psi_a'(x') = (x^2)^{\Delta - 1/2} x \cdott \gm^b_a
\psi_b(x') \; , \label{confinverse}
\end{equation}
where ${x_i}' = +x_i / x^2$ and $a,b$ are spinor indices. We see from
(\ref{fuzzyreps}) that the global Hilbert space now always splits into
two subspaces which are dual to each other under (\ref{confinverse}).
This map is an inversion around the equator. The pure monomial
(i.e. not involving $x\cdott\gamma$) 
with high degree are concentrated closer to the horizon. We interpret
therefore the pure monomials with highest 
degree and their mirror images with a factor $x\cdott\gamma$ as the
states near the horizon. As we will discuss, the number of these states also grow as the area of the horizon.

To find the coordinate representations of an orthonormal basis for an
observer's Hilbert space, we should again diagonalize the $L_0$
eigenbasis, (\ref{fuzzyreps}), with respect to the Cartan subalgebra
of $so(d-1)$ and its Casimir. The latter is now
\begin{equation}
J^2 = L^2 + x \cdot \pl - x \cdott \gm ~ \gm \cdott \pl + {1 \over 8}
(d-1)(d-2) \; ,
\end{equation}
which one can check reduces to $(L + S)^2$ for $d = 4$. In $d=4$, one can use
the eigenfunctions of $L_3$ and $L^2$ from the previous subsection to
construct eigenfunctions of $L_0$, $J_3$, and $J^2$. Denoting the coordinate
representation of $| h \, j \, m_j \ket$ as $\psi_{h, j, m_j}$, a short
exercise in computing Clebsch-Gordan coefficients gives, for example:
\begin{equation}
\psi_{\Delta,\frac{1}{2},\frac{1}{2}} (x)
=  N_{1/2} \lf \begin{array}{cc}1 \\ 0 \end{array} \rt \comma
\psi_{\Delta,\frac{1}{2},-\frac{1}{2}} (x)
=  N_{1/2} \lf \begin{array}{cc}0 \\ 1 \end{array} \rt \; ,
\end{equation}
which both have $L_0$ eigenvalue $+\Delta$ and $J^2$ eigenvalue 3/4, and
\begin{eqnarray}
\psi_{\Delta-1,\frac{3}{2},\frac{3}{2}} (x)
=  N_{3/2} \lf \begin{array}{cc}{1 \over \sqrt{2}} (x_1 +
  i x_2) \\ 0 \end{array} \rt \comma
\psi_{\Delta-1,\frac{3}{2},\frac{1}{2}} (x)
=  N_{3/2} \lf \begin{array}{cc} - \sqrt{2} x_3 \\ 
 {1 \over \sqrt{2}} (x_1 + ix_2) \end{array} \rt \nonumber \\
\psi_{\Delta-1,\frac{3}{2},-\frac{1}{2}} (x)
 =  N_{3/2} \lf \begin{array}{cc} 
 -{1 \over \sqrt{2}} (x_1 - ix_2) \\ - \sqrt{2} x_3  \end{array} \rt \comma
\psi_{\Delta-1,\frac{3}{2},-\frac{3}{2}} (x)
 =  N_{3/2} \lf \begin{array}{cc} 0 \\ {1 \over \sqrt{2}} (x_1 -
  i x_2) \end{array} \rt \; ,
\end{eqnarray}
which both have $L_0$ eigenvalue $+\Delta-1$ and $J^2$ eigenvalue 15/4.

A positive-definite hermitian inner product for the spinor fields is
\begin{equation}
(\psi_m, \psi_n) = \int {d^{d-1} x \over (1+x^2)^{d-3/2+\Delta}} \lf
  \psi^*_m(x) x \cdott \gm \psi'_n(x') + \psi_n(x) x \cdott \gm^*
  \psi^{'*}_m(x') \rt \; . 
\end{equation}
This inner product is $O(d-1)$-invariant.

\subsection{Counting of states and the large representation limit}

To find the size of these representations, we need to count the number of
symmetric monomials as well as the coordinates with $x\cdott \gm$.
In addition, there is a spinor index attached to everything. So for
example, when $d=4$, there are 4 global Hilbert states at $\Delta =
1/2$, 16 at $\Delta = 3/2$, and 40 at $\Delta = 5/2$ etc. Note that the
$\Delta = 1/2$ representation is just the spinor representation of the
previous subsection, the four states corresponding to spin-up and
spin-down for either antipodal observer. The antipodal states can be
viewed as being created by spinor fields acting at $1$ and at
its inverse, $(x^2)^\Delta\, x \cdott \gm$. The dimension of the global
Hilbert space in representation $\Delta$ for $d$ even is
\begin{equation}
{\rm dim}~ h_\Delta = 2 \cdot 2^{(d-2)/2} \sum_{p=0}^{\Delta-1/2} 
\lf \begin{array}{c} d + p - 2 \\ p \earr \rt = 
2^{d/2} \lf \begin{array}{c} d + \Delta - 3/2 \\ \Delta - 1/2 \earr \rt \;
  . \label{fuzzyspinrepsize}
\end{equation}
For odd $d$, we just replace $2^{(d-2)/2}$ with $2^{(d-1)/2}$. 

The global Hilbert states are labeled
by an $O(1,d)$ index coming from the spinor, as well as an $O(1,d)$
index coming from the coordinates. This leads to a further
decomposition as, for example,
\begin{equation}
{\bf 4} \otimes {\bf 14} = {\bf 40} \oplus {\bf 16} \; ,
\end{equation}
so an $O(1,4)$ spinor (the {\bf 4}) on the 14 coordinates of
(\ref{Delta=2}) is reducible into a $\Delta = 3/2$ and a $\Delta = 5/2$
representation. Again the fact that these representations are finite
is indicative of an underlying exclusion principle. We have
essentially discretized the sphere. Indeed, some of the expressions
are reminiscent of quantum foam \cite{foam}.

Consider now what happens as the cosmological constant is decreased.
As the horizon area increases, the number of states also
increases so we are going towards larger representations. But now the
size of the finite conformal representation on the sphere is related
to the radius of the sphere itself (which is a Casimir of the
representation). We see from (\ref{fuzzyspinrepsize}), that the
dimension of the tensor product space is
\begin{equation}
\ln {\rm dim}~ H_{\Delta} = \lf {\rm dim}~ h_{\Delta} \rt \ln 2
= \left [ 2^{d/2} \lf \begin{array}{c} N + d - 1\\
      N \earr \rt  \right ] \ln 2 \; ,
\end{equation}
where $N = \Delta - 1/2$ and $d$ has been taken to be even. The
logarithm of the number of states in the Fock space therefore grows as
\begin{equation}
{1 \over 2} \ln {\rm dim}~ H \sim N^{d-1} \; .
\end{equation}
We see that the logarithm of the number of Fock states scales as the
volume of the boundary sphere. These are states that live in one
hemisphere of the boundary sphere. We can obtain an answer that scales as
the area, by considering only those states that are entangled with
states in the other hemisphere, i.e. with the states of the antipodal
observer. These are the states at the equator of the sphere:
\begin{equation}
S = \left [ 2^{d/2} \lf \begin{array}{c} N + d - 2 \\ N \earr \rt
  \right ] \ln 2 \sim N^{d-2} \; .
\end{equation}
Hence our construction seems to suggest that the entropy of de
Sitter space counts the number of states entangled across the horizon
\cite{dangilad}.

Another interesting issue is the restoration of unitarity. 
The tensor product states form finite-dimensional representations of
$O(1,d)$. Obviously, these representations are nonunitary. We can
write a global Hilbert state as
\begin{equation}
\Psi = \psi(x) + \psi'(x') = \psi(x) + x \cdott \gm (x^2)^{\Delta -
  1/2} \psi(x') \; .
\end{equation}
Then an $O(1,d)$-invariant hermitian bilinear is
\begin{equation}
(\Psi_1, \Psi_2) = \int {d^{d-1} x \over (1+x^2)^{d-1+\Delta}} 
  \Psi^\dagger_1(x) \Gamma^0 \Psi'_2(x) \; .
\end{equation}
This does not qualify as an inner product since it is not
positive-definite; half the states have negative norm.
Indeed, if one thinks of $h_I$ as being the electron
Hilbert space and $h_{II}$ as being a positron Hilbert space, then
$\overline{\Psi} \Psi$ is just $(e^\dagger e - p^\dagger p)$. 
From the inner product, and the way the generators act, it is clear that
the nonunitarity of the representations is associated with the
existence of the other side of the horizon. Furthermore, the
construction is symmetric under the exchange of positive and negative
norm states, very much like the symmetry between negative and positive
energy states for Dirac spinors. This suggests that we should try to
take the analogy further and ``fill'' the other side of the horizon like
a Dirac sea. This leads to the appealing picture that the horizon
becomes identified with a Fermi surface, an idea that has started to
appear in other contexts as well. Furthermore, the states right at the
surface grow precisely as the area of the horizon. 

Consider now the generators $L_{i\pm}$. These generators can mix
states of different observers by moving monomials across the
equator. On the complete Fock space  this action is
nonunitary, since a positive norm state for one observer is
in general a mixture of positive and negative norm states for
another.\footnote{This phenomenon is closely related to the Bogolubov
transformation that is responsible for Hawking
radiation in de Sitter space.} But note that only those states that are
already at the equator are moved into the antipodal observer's
states. States at other positions are merely moved closer or farther
from the equator; on these states, the operators $L_{i \pm}$ are
hermitian. For a given $N$, the ratio of equatorial states to  total
states falls as $1/N$. So, for very large representations, the
conformal generators act in an increasingly unitary manner on states
that belong to a single observer's Fock space. In the Minkowski limit
of an infinitely large representation, the generators become precisely
unitary. Each Fock space then furnishes a unitary infinite-dimensional
representation of the Poincar\'e group, and the two Fock spaces of the
pair of antipodal observers decouple. The vanishing cosmological
constant limit of de Sitter space is thus {\em two} copies of
Minkowski space.

\section{De Sitter-Invariance and the S-matrix}

The spinor fields on the sphere give an infinite class of
finite-dimensional nonunitary representations
of the de Sitter group, all of which can be written as $h_I\oplus
h_{II}$, with $h_I$ and $h_{II}$ both being 
unitary representations of the rotation group. 
We interpret these as the
one-particle Hilbert space for a given observer. The Fock spaces $H_I$
and $H_{II}$ can thus be 
constructed as being two isomorphic but otherwise independent spaces. 
As described, these Hilbert space are representations of the little group that 
leave the horizon invariant. One can think of the corresponding charges as
being defined at the horizon (note that, when going to the Minkowski
space limit it becomes asymptotic infinity). In global de
Sitter space there is no
boundary, and hence no conserved charges. Therefore
global states correspond to de Sitter-invariant tensor product states. 

Observer complementarity states that
each observer has complete but complementary information. The global states in
de Sitter space do not give complete information to single observers, since
they are described as entangled states. We can implement observer
complementarity by making use of an antipodal map. As we will
see, if we regard $H_I$ as the space of in-states, the antipodal
identification takes $H_{II}$ and identifies it with the space of
out-states. Furthermore, the de Sitter-invariant global state becomes
the S-matrix that maps in-states to out-states. De Sitter
invariance of the S-matrix is very natural: just as in Minkowski space, we
want the S-matrix to respect the spacetime isometry group.

\subsection{Antipodal map}

The elliptic interpretation of de Sitter space
\cite{Schroedinger,dSZ2,Gibbons,Sanchez}
consists of identifying
events that are related by an involution, the $\ZZ$ antipodal map
\begin{equation}
X^I\to -X^I \; ,
\end{equation}
where $I=0,1,\ldots, D$, together with charge conjugation, C
\cite{dSZ2}. Here $X^I$ are the Cartesian embedding coordinates of
$d+1$-dimensional Minkowski space; see (\ref{embed}). The fixed point
of this identification, $X^I = 0$ is not itself on the de Sitter hyperboloid,
so this is a freely-acting symmetry. The quotient space, $dS/\ZZ$ or
``elliptic de Sitter space,'' is therefore a homogeneous space with no
special points. Moreover, the transformation $X^I \to -X^I$ is in the
center of the de Sitter group so the quotient space still has the same
local symmetries and the question of finding finite-dimensional
representations of the de Sitter group continues to apply. 
For a local observer, the geometry is unchanged but 
antipodal points on the horizon are identified. Indeed, if we think of
de Sitter space as the Lorentzian version of a sphere, then elliptic
de Sitter space is the Lorentzian version of a projective sphere. 

\begin{figure}[hbtp]
        \centering
\begin{picture}(0,140)
\Oval(0,132.5)(15,60)(0)
\Oval(0,7.5)(15,60)(0)
\CArc(-167,70)(122.71,-29,29)
\CArc(167,70)(122.71,151,209)
\ArrowLine(-44.29,70.01)(-44.29,69.99)
\ArrowLine(44.29,69.99)(44.29,70.01)
\put(32,26){\ding{72}} 
\LongArrow(33,34)(28,42)
\SetColor{Gray}
\Line(0,7.5)(0,132.5)
\put(-35,104){\ding{73}} 
\LongArrow(-28,101)(-23,93)
\SetColor{Black}
\end{picture}
\caption[]{\small The antipodal map reverses the local arrow of time.}
      \label{fig:antipode}
\end{figure}
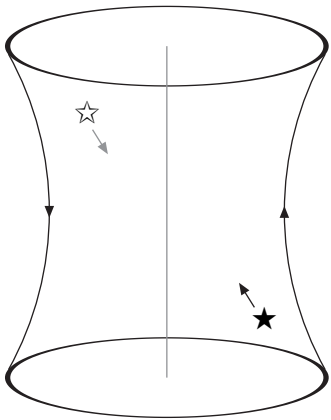

Note that the antipodal map also inverts the direction of time; see
Figure \ref{fig:antipode}. As a result, elliptic de Sitter space is not
time-orientable; there is no globally-consistent way to distinguish
future light cones from past light cones.
For example, consider global coordinates. The line element reads
\begin{equation}
ds^2=-dT^2+R^2\cosh^2 (T/ R) \lf d\theta^2 + \sin^2\theta~
d\Omega^2_{D-2} \rt \; .
\end{equation}
In these coordinates the antipodal map is given by
\begin{equation}
T\to-T\qquad \theta\to\pi-\theta\qquad \Omega\to \Omega^A \; ,
\end{equation}
where $\Omega^A$ are the angular coordinates of the point antipodal on the
$D-2$-dimensional sphere to the point labeled by $\Omega$, and time is
reversed, $T \to -T$. One can check that these identifications, even
though they involve time, do not lead to any obvious inconsistencies
involving causality \cite{Schroedinger,dSZ2} such as closed timelike curves.
There are several interesting consequences of making such an
identification. Since the spacetime has effectively been halved by the
$\mathbb{Z}_2$ identification, every observer now has complete
information, in the sense that there are no independent events outside
his or her horizon. Moreover, the global spacetime now has only one
asymptotic boundary, rather than two, since $\scriplus$ and
$\scriminus$ have been identified. This seems holographically more
appealing since otherwise the dual theory would live on two
disconnected manifolds.

Now the arrow of time in the antipodal observer's causal patch points in
the opposite sense. This suggests that if an observer's space can be
thought of as the space of initial states, then the antipodal
Fock space should be regarded as the {\em final} state
space \cite{dSZ2,gerard}. The tensor product states $|m \ket_I \otimes
|n \ket_{II}$ should now be thought of as a physical process $|m
\ket_{\rm in} \otimes |n' \ket_{\rm final}$ where the prime indicates
that the final state is a CPT conjugate of the corresponding
ket e.g. $\up' = \down$. We also have a choice, whether to
map $|0 \ket_{\rm in}$ to $\bra 0|_{\rm out}$ or to $\bra \,
\uparrow \downarrow \! |_{\rm out}$. Then taking the hermitian
conjugate of (\ref{psi}), we find
\begin{equation}
\Psi = \sum_{m,n} C^*_{mn} | n' \ket_{\rm out} \otimes 
\bra m|_{\rm in} \; , 
\end{equation}
and hence $C^\dagger$ is just the S-matrix. The invariant tensor
product states (\ref{invariants}) are then the de Sitter-invariant
building blocks of the S-matrix.

\subsection{An illustration in terms of spinors}

As an illustration let us find the invariant tensor product for the
Dirac spinor construction in $d=4$; the same arguments will hold for
the more general representations corresponding to the spinor field on the
sphere. Our one-particle Hilbert space is therefore just $\up$ and
$\down$; the Fock space is spanned by the four states $|0 \ket$,
$\up$, $\down$, and $\ud$. These four states can be labeled by their
transformation properties under $O(3)$. Obviously, they belong to the
$\bf{1}$, the $\bf{2}$, and the $\bf{1}$ representations, respectively. 

The antipodal observer has an isomorphic Fock space. The tensor
product states therefore transform under the direct product of the
respective representations. Taking the direct product of the $\bf{1}$,
the $\bf{2}$, and the $\bf{1}$ with themselves we find that the 16 tensor
product states belong to
\begin{equation}
{\bf 1} \oplus {\bf 2} \oplus {\bf 1} \oplus 
{\bf 2} \oplus {\bf 3} \oplus {\bf 1} \oplus {\bf 2} \oplus
{\bf 1} \oplus {\bf 2} \oplus {\bf 1} \; .
\end{equation}
These are $O(3)$ labels. We are interested in how the
tensor product states transform not just under rotations, but under
general de Sitter transformations. We can ``recompose'' the above
$O(3)$ representations into $O(1,4)$ representations:
\begin{equation}
{\bf 1} \oplus {\bf 4} \oplus {\bf 5} \oplus 
{\bf 1} \oplus {\bf 4} \oplus {\bf 1} \; . \label{so14reps}
\end{equation}
For example, the states $|0\ket_I \otimes \up_{II}$, 
$|0\ket_I \otimes \down_{II}$, 
$\up_I \otimes | 0 \ket_{II}$, and
$\down \otimes |0 \ket_{II}$ combine to form a $\bf{4}$, a spinor of
$O(1,4)$. The $\bf{5}$ is a vector of $O(1,4)$. 

Notice, in particular, that there are three singlet states. These are
tensor product states that are invariant under the action of the de
Sitter group. The three invariant states are
\begin{equation}
|0\ket \otimes |0 \ket \qquad \ud \otimes \ud \qquad \lf
\up \otimes \down - \down \otimes \up \rt \; .
\label{invariants}
\end{equation}
As we have discussed, under the antipodal identification, tensor product states
correspond to physical processes: the states in Fock space $I$
are initial states while the states in Fock space $II$ become final states.
There are therefore three de Sitter-invariant processes in this simple model.

 With $|0 \ket_{\rm in}
\to \bra \, \uparrow \downarrow \! |_{\rm out}$, and with the rows and
columns labeled in the order $|0 \ket$, $\up$, $\down$, $\ud$, a de
Sitter-invariant S-matrix is
\begin{equation}
S = \lf \barr 
0 \; \; \; 0 \; \; \; 0 \; \; \, a \\
0 \; \; \; b \; \; \; 0 \; \; \; 0 \\
0 \; \; \; 0 \, \; \; b \; \; \; 0 \\
c \; \; \; 0 \; \; \; 0 \; \; \; 0 
\earr \rt \; ,
\end{equation}
where we have used $T \up = \eta \down$ and $T \down = - \eta \up$
for time-reversal on spinors. If $a$, $b$, and $c$ are all phases,
then the S-matrix is unitary as well as de Sitter-invariant. We learn
that the number of independent S-matrix elements is the number of
invariant tensor product states.

More generally, we can consider a Fock space that is the product of $N$
different Fock spaces, each the Fock space of a different species of
spinor. One may think of $N$ as being the number of flavors. 
A basis state for this multi-species Fock space is
\begin{equation}
| \chi \ket_1 \otimes | \chi \ket_2 \otimes \ldots \otimes | \chi
  \ket_N \; ,
\end{equation}
where each $| \chi \ket$ can be $|0 \ket$, $\up$, $\down$, or
$\ud$. For $N$ spinors there are $4^N$ Fock states and $4^{2N}$ tensor
product states.
Consider for example $N = 2$. An individual observer's Fock space is
now itself a direct product of the Fock spaces of the two spinors. The
$O(1,4)$ labels for the representations that the 16 Fock space states
transform under are given precisely by (\ref{so14reps}). The same set
of representations applies to the antipodal observer's Fock space.
There are then 256 tensor product states describing global de Sitter
space. They are grouped in representations according to the direct
product of $O(1,4)$ representations. We use the fact that
\begin{eqnarray}
{\bf 4} \otimes {\bf 4} 
& = & {\bf 10} \oplus {\bf 5} \oplus {\bf 1} \nonumber \\
{\bf 4} \otimes {\bf 5}
& = & {\bf 16} \oplus {\bf 4} \nonumber \\
{\bf 5} \otimes {\bf 5} 
& = & {\bf 14} \oplus {\bf 10} \oplus {\bf 1} \; .
\end{eqnarray} 
Hence the 256 tensor product states transform under
\begin{equation}
14 \times {\bf 1} \oplus 16 \times {\bf 4} \oplus 10 \times {\bf 5} \oplus 
5 \times {\bf 10} \oplus 4 \times {\bf 16} \oplus 1 \times {\bf 14} \; .
\end{equation}
Here there are 14 de Sitter-invariant tensor product states. 
This means that a de Sitter-invariant S-matrix can at most depend on
14 independent parameters. 

Just as an aside, we note that the ${\bf 1}$, ${\bf 4}$, ${\bf 5}$,
${\bf 10}$, ${\bf 16}$, and the ${\bf 14}$ have a familiar
interpretation if we regard $O(1,4)$ as the Lorentz group in five
dimensions: in
this language they represent the scalar, spinor, vector, antisymmetric
tensor, gravitino, and graviton, respectively and form the ${\cal N} =
2$ supergravity multiplet. This is perhaps not surprising since we
built the representation out of two independent spinors. Notice
further that the multiplicities are themselves the size of
representations of a five-dimensional orthogonal group, which reflects
the fact that in this toy example the symmetry group is actually $SO(5) \times
SO(1,4)$.

The number of de Sitter invariants is in general much less
than the number of conditions imposed by unitarity:  in our example there are 256
S-matrix elements but only 14 invariants. In order for the S-matrix to
be unitary, these 14 complex parameters have to be chosen to satisfy
the 128 complex equations imposed by unitarity. The requirements of de
Sitter symmetry and unitarity are therefore very constraining. One
may worry that there may not be any solution but, as one can
explicitly check, it is possible to satisfy de Sitter invariance and
unitarity and still have a small but nonzero set of free parameters.
This check has been preformed for small representations, and works remarkably.
We do not have a general proof, but we regard the fact that is works in these case 
as an indication that at least in principle their is no clash between de Sitter invariance 
and unitarity. In fact, by considering larger representations and allowing several spinor fields we expect that it even becomes easier to obey both requirements. Finally, we note that gauge invariance can also be included in our 
construction by giving the spinor fields an extra representation index and requiring that the initial and final states are gauge invariant.  

\bigskip

\section{Observer Complementarity}

We have argued that, after making the antipodal identification, every
observer has complete information, and tensor product states become
processes. However, the physical interpretation of these processes
generally depends on the observer. This is a manifestation of what we
call ``observer complementarity,'' the notion that each observer can,
in principle, describe everything that happens within his or her
horizon using only pure states.  The
physics however may appear in rather different -- complementary -- guises.  

Our construction gives a realization of this notion of 
observer complementarity, not just in terms of pictures or words, but in
the actual expressions. 
Consider the
simplest case of one spinor in $d = 4$. We saw that there were only
three de Sitter-invariant states, (\ref{invariants}). For example, the
process
\begin{equation}
\left \{ \up \to \down \right \} - \left \{ \down \to \up \right \}
\end{equation}
corresponds to a spin-flip, and all observers can agree on
that. In contrast, spin-up going to spin-down is {\em not} a process that
all observers agree on because
\begin{equation}
\up \to \down = {1 \over 2} \lf \left \{ \up \to \down \right \} -
\left \{ \down \to \up \right \} \rt
+ {1 \over 2} \lf \left \{ \up \to \down \right \} + \left \{ \down
\to \up \right \} \rt \; ,
\end{equation}
and only the first term in parentheses transforms in the singlet; the
second term transforms in the ${\bf 5}$ of $O(1,4)$.

More generally, the S-matrix contains the probability amplitudes for
all possible events or processes. Every event is visible to all observers.
But different observers give different interpretations of the same event
in terms of a scattering process of in- to out-states. This is because
the de Sitter group element that relates two observers mixes up the in- and
out-states. At first this may seem to be in conflict with de Sitter
invariance, but it is not. De Sitter invariance implies only that the
probability amplitudes that each observer associates with the observed
event are the same. The physical interpretation of the event as a
scattering process may well be different. This is just like in 
Minkowski space, where generally in- and out-states do transform under
the Lorentz group (though of course they do not mix), but may be
expressed in terms of Lorentz-invariant ``form factors.'' In this sense,
the number of de Sitter-invariant tensor products counts how many form
factors can appear in the most general scattering process.

\bigskip
\noindent
{\bf Acknowledgments}

\noindent
We like to thank Bartomeu Fiol, Lenny Susskind and Herman Verlinde for discussions.
M.~P. is a Columbia University Frontiers of Science fellow and is
supported in part by DOE grant DF-FCO2-94ER40818.

\end{document}